\documentclass[amsmath,amssymb,floatfix,preprint,showpacs,superscriptaddress]{revtex4}

\usepackage{graphicx}
\usepackage{dcolumn}
\usepackage{bm}

\begin{document}


\title{A parsimonious model for intraday European option pricing}

\author{Enrico Scalas}
\email{scalas@unipmn.it}
\homepage{www.mfn.unipmn.it/~scalas}
\affiliation{BCAM - Basque Center for Applied Mathematics, Bilbao, Basque Country, Spain and 
Dipartimento di Scienze e Tecnologie Avanzate, Universit\`a del
Piemonte Orientale ``Amedeo Avogadro'', Alessandria,
Italy}

\author{Mauro Politi}
\affiliation{BCAM - Basque Center for Applied Mathematics, Bilbao, Basque Country, Spain}

\date{\today}



\begin{abstract}
A stochastic model for pure-jump diffusion (the compound renewal process) can be used as a 
zero-order approximation and as a phenomenological description
of tick-by-tick price fluctuations. This leads to an exact and explicit general formula for the martingale price of a 
European call option. A complete derivation of this result is presented by means of elementary probabilistic tools. 
\end{abstract}

\maketitle

\section{Introduction}

Market microstructure and high-frequency trading are fields experiencing an 
increase of interest within financial institutions and academic scholars (see \cite{foresight11}).
Therefore, models for tick-by-tick financial fluctuations, that might have been considered just 
a curiosity more than a decade ago (see the seminal paper by \cite{engle98}), are becoming more and more important
for practical purposes (see \cite{scalas00} and
\cite{mainardi00} for early studies of the model presented below). The literature on high-frequency price modelling
up to the middle of the last decade is reviewed in \cite{hautsch04} and in \cite{kokot04}.

Along with interest in high-frequency trading comes the need for high-frequency hedging. In this work,
we address the problem of determining the price for an intra-day European option
written on a share traded in a stock exchange.
It is assumed that the derivative position is opened at a time $t$ after the start of continuous
trading with maturity at a time $T_M$ before the end of continuous trading on the very same day. The fluctuations
of the share price $S(t)$ can be modeled by a suitable c\`adl\`ag (i.e. right-continuous
with left limit) pure-jump process. One might be tempted to use a compound Poisson process
for the logarithm of the price $X(t) = \log(S(t)/S(0))$ and immediately apply the result 
of \cite{merton1976} with the coefficients of drift and diffusion set to zero. For a
vanishing risk-free interest rate (which is a reasonable assumption for intra-day data, see the discussion below),
this would lead to the following formula for the plain-vanilla option price $C(t)$
\begin{equation}
\label{mertonformula}
C(t) = \mathrm{e}^{-\lambda (T_M -t)} \sum_{n=0}^\infty \frac{(\lambda (T_M -t))^n}{n!} C_n (S(0),K,\mu,\sigma^2),
\end{equation}
where $\lambda$ is the activity of the Poisson process for trades, $K$ is the strike price, $\mu$ and
$\sigma^2$ are, respectively, the expected value and the variance of the log-price jumps which are assumed to be
normally distributed. One further has that
\begin{equation}
\label{discreteoptionprice}
C_n (S(0),K,\mu,\sigma^2) = N(d_{1,n}) S(0) - N(d_{2,n}) K, 
\end{equation}
where
\begin{equation}
\label{standnorm}
N(u) = \frac{1}{\sqrt{2 \pi}} \int_{-\infty}^u dv \mathrm{e}^{-v^2/2}
\end{equation}
is the standard normal cumulative distribution function and, finally
\begin{equation}
\label{d1}
d_{1,n} = \frac{\log(S(0)/K) + n (\mu + \sigma^2/2)}{\sqrt{n} \sigma},
\end{equation}
\begin{equation}
\label{d2}
d_{2,n} = d_{1,n} - \sigma \sqrt{n}.
\end{equation}

Whereas Merton's jump diffusion model is still the object of active research and is discussed
in a recent research paper by \cite{cheang11}, it has several unrealistic
features. One of them is that unconditional inter-trade durations
do not follow the exponential distribution 
(see \cite{engle97, engle98, mainardi00, raberto02, scalas04a, scalas06}). Semi-Markov models
(described in \cite{janssen07}) can take this fact into account as shown by 
\cite{scalas04a} and \cite{scalas11a}. In the following, we shall consider 
intra-day European options written on semi-Markov pure jump models which are compound renewal
processes (see also
\cite{scalas11a} and \cite{baleanu12}). Related papers are \cite{montero08} and \cite{cartea10}.
In \cite{montero08}, the focus is on option prices for derivatives written on compound Poisson
processes and in the presence on non-vanishing risk-free interest rate, whereas \cite{cartea10}
extends L\'evy option prices to the semi-Markov case by developing suitable approximations.
Finally, a recent paper by \cite{shaw11} considers Laplace transform methods to deal with order
and trade renewal flows in an agent-based model where the trade counting process is not
necessarily Poisson. 

The model proposed here has several distinctive advantages; 
for instance, each model entity has an immediate and clear translation into a microstructural 
quantity (there are no hidden variables of any kind); moreover, it is very parsimonious (leading 
to a many sound mathematical results); finally, the model is very flexible.

In what follows, section \ref{tick} will present the model, section \ref{option} will be devoted to pricing 
European options, and section \ref{concl} will report a discussion of the results. We would like to stress 
that the derivations and computations below will involve only elementary probabilistic methods.

\section{An Elementary but Comprehensive Model for Tick-by-tick Price Fluctuations}
\label{tick}

Let $S(t)$ denote the price of an asset at time $t$ and let $X(t) = \log (S(t)/S(0))$ be
the corresponding logarithmic price (or log-price), where $S(0) = S(t=0)$ will be assumed 
to be $S(0) = 1$ without loss of generality. We can take as $S(0)$ the opening
price of the asset after the opening auction and before the beginning of continuous trading
in a stock market. After continuous trading begins, trades will take place at specific
{\em epochs} $\{T_i\}_{i=1}^M$ where $M$ denotes the total number of trades within the
day. We shall further assume that $T_0 = 0$.
The trading epochs can be seen as a {\em point process} (see \cite{daley88}).
Our first assumption is that these epochs are a {\em renewal process} (see \cite{cox70}),
meaning that the
{\em inter-trade durations} $J_i = T_i - T_{i-1}$
are independent and identically distributed 
(i.i.d.) positive random variables.
Notice that, for the $n$-th epoch, one has
\begin{equation}
\label{time}
T_n =\sum_{i=1}^n J_i,
\end{equation}
even if the i.i.d. hypothesis is not satisfied. If the durations $\{J_i\}_{i=1}^{\infty}$
are i.i.d. random variables, there is a simple and convenient relationship between the distribution
of durations and the distribution of epochs. Indeed, given two independent random variables
$U$ and $V$ with respective cumulative distribution functions $F_U (u) = \mathbb{P}(U \leq u)$ and
$F_V (v) = \mathbb{P} (V \leq v)$, one can compute the cumulative distribution function $F_W(w)$ of their
sum $W = U + V$, which turns out to be the the measure convolution (a.k.a. Lebesgue-Stieltjes convolution) of the 
two distributions for $U$ and $V$ (see \cite{bingham87}). 
Throughout this paper, we shall use the so-callled {\em indicator-function method}, 
a procedure allowing to write probabilities as expectations. 
Therefore, it is instructive to show in detail how the distribution of the sum of two independent
random variables can be derived, even if this is a well-know result. The first step is to notice that the joint
cumulative distribution function $F_{U,V} (u,v)$ is given by $F_{U,V} (u,v) = F_U (u) F_V (v)$ as a consequence of
independence. The second step is to recall that the probability of an event $A$ is given by the expected value of
the indicator function $I_A$, namely $\mathbb{P} (A) = \mathbb{E} (I_A)$ and 
that the indicator function of the intersection of two events $A$ and $B$ is the product of the 
parent indicator functions, i.e. we have $I_{A \cap B} = I_A I_B$. The last step is to notice that the 
event $\{ W \leq w\}$ is equivalent to $\{ U \in \mathbb{R} \} \cap \{ V \leq w - U \}$.
Therefore, one has the following chain of equalities
\begin{eqnarray}
\label{proof1}
F_W(w) & = & \mathbb{P} (W \leq w) = \mathbb{E} \left( I_{ \{W \leq w \} } \right) = E \left( I_{\{U \in \mathbb{R}\}} 
I_{\{V \leq w - U\}} \right) \nonumber \\
& = & \int_{u \in \mathbb{R}} \int_{v \leq w - u} dF_{U,V} (u,v) = \int_{u \in \mathbb{R}} \int_{v \leq w - u} dF_U(u) dF_V(v) \nonumber \\
& = & \int_{u \in \mathbb{R}} dF_U (u) \int_{v \leq w - u} dF_V (v) = \int_{u \in \mathbb{R}} F_V(w-u) dF_U (u).
\end{eqnarray}
To denote the convolution, which is an operation symmetric in $U$ and $V$, we can introduce the symbol $\star$:
\begin{eqnarray}
\label{convolutionsymbol}
F_W(w) & = & \int_{u \in \mathbb{R}} F_V(w-u) dF_U (u) = \int_{v \in \mathbb{R}} F_U(w-v) dF_V (v) \nonumber \\
& = & F_U \star F_V (w) = F_V \star F_U (w).
\end{eqnarray}
This formula holds true also if $U$ and $V$ are positive random variables. In such a case one has that
$F_U (u) = 0$ for $u < 0$ and that $F_V (v) = 0$ for $v <0$ (notice that $F_U (0)$ and $F_V (0)$
may be positive). Then equation \eqref{convolutionsymbol} becomes 
\begin{equation}
\label{convpos}
F_W(w) = \int_{0}^w F_V(w-u) dF_U (u) = \int_{0}^w F_U(w-v) dF_V (v).
\end{equation}
Now, let $F_J (x)$ denote the cumulative distribution
function of the duration, i.e. $F_J (t) = \mathbb{P} (J \leq t)$; moreover, let $F_{T_n} (t)$
denote the cumulative distribution function of the $n$-th epoch, i.e. $F_{T_n} (t) = \mathbb{P} (T_n \leq t)$.
Then, $F_{T_n} (t)$ is given by the $n$-fold convolution of $F_J (t)$, that is by
\begin{equation}
\label{measureconvolution}
F_{T_n} (t ) = F^{\star n}_J (t).
\end{equation}
Equation \eqref{measureconvolution} can be proved by means of the iterated application of equation \eqref{convpos}.

A price $S(T_i)$ corresponds to each trading epoch $T_i$. Let $Y_i = \log(S(T_i)/S(T_{i-1}))$ represent
the {\em tick-by-tick logarithmic return}, then the log-price $X(t)$ is given by
\begin{equation}
\label{logprice}
X(t) = \sum_{i=1}^{N(t)} Y_i,
\end{equation}
where the {\em counting process} $N(t)$ is defined as
\begin{equation}
\label{countingprocess}
N(t) = \mathrm{max} \{n:\, T_n \leq t\},
\end{equation}
and counts the number of trades since the beginning of continuous trading. The relationship between the
log-price and the price is
\begin{equation}
\label{priceequation}
S(t) = \mathrm{e}^{X(t)} = \mathrm{e}^{\sum_{i=1}^{N(t)} Y_i} = \prod_{i=1}^{N(t)} \mathrm{e}^{Y_i}.
\end{equation}
We shall further assume that $\{Y_i\}_{i=1}^{N(t)}$ is a sequence of i.i.d.
random variables such that $\bar{Y} = \mathbb{E}(Y_i) < \infty$.
Let $\mathcal{F}_t$ denote the natural filtration of the process $S(t)$ up to time $t$, this being the $\sigma$-field
generated by the random variables $T_1, \ldots, T_{N(t)}$ and $Y_1, \ldots, Y_{N(t)}$.

In general, with the above hypotheses, $S(t)$ given by equation \eqref{priceequation} is not a martingale. In fact
one has for $s < t$
\begin{eqnarray}
\label{martingale}
\mathbb{E} (S(t)|\mathcal{F}_s) & = & \mathbb{E} \left( \prod_{i=1}^{N(t)} \mathrm{e}^Y_i| \mathcal{F}_s \right) 
= \prod_{i=1}^{N(s)} \mathrm{e}^{Y_i} \mathbb{E} \left( \prod_{i=N(s)+1}^{N(t)} \mathrm{e}^Y_i| \mathcal{F}_s \right) \nonumber \\
& = & S(s) \prod_{i= N(s) + 1}^{N(t)} \mathbb{E} \left( \mathrm{e}^{Y_i} \right),
\end{eqnarray}
and the martingale condition is statisfied only if for every $s, t$ such that $s < t$, one has
\begin{equation}
\label{martingalecondition}
\prod_{i= N(s) + 1}^{N(t)} \mathbb{E} \left( \mathrm{e}^{Y_i} \right) = 1;
\end{equation}
this is the case if $\mathbb{E} ( \mathrm{e}^{Y_i} ) =1$. 
However, one can always find an equivalent martingale measure (e.m.m.).
One can replace $Y_i$ in equation \eqref{logprice} with $Y_i - a$ defining the following processes, a modified log-price process
\begin{equation}
\label{logpricemart}
\widetilde{X} (t) = \sum_{i=1}^{N(t)} (Y_i - a),
\end{equation}
as well as the corresponding modified price process
\begin{equation}
\label{pricemart}
\widetilde{S} (t) = \mathrm{e}^{\widetilde{X}(t)}.
\end{equation}
Now, if $a = \log(\mathbb{E}(\mathrm{e}^{Y_i}))$, one has that $\widetilde{S} (t)$ is a martingale. In fact, one can write
\begin{equation}
\label{martproof}
\mathbb{E} (\widetilde{S}(t)|\mathcal{F}_s) = \widetilde{S}(s)  \prod_{i= N(s) + 1}^{N(t)} \mathbb{E} \left( \mathrm{e}^{Y_i -a} \right)
= \widetilde{S} (s).
\end{equation}
Armed with this e.m.m., it is possible to move on and price options written on the process defined above using the martingale method.

\section{Martingale Option Pricing}
\label{option}

For an intra-day time horizon, we can safely assume that the 
{\em risk-free} interest rate is zero; even if such a return rate 
were $r_Y = 10 \%$ on a yearly time horizon, meaning that the institution or government  issuing this instrument is close to default (so that,
it would not be so riskless, after all) or that
the inflation rate is quite high, the interest rate for one day would be $r_d \approx 1/(10 \cdot 200) = 5 \cdot 10^{-4}$ ($200$
is the typical number of working days in a year) and
this number has still to be divided by $8$ (number of trading hours) and then by $3600$, if the goal is 
approximating the rate at the time scale of one second. This gives $r_s \approx 1.7 \cdot 10^{-8}$.
On the other hand, typical tick-by-tick returns in a stock exchange are larger than the tick divided by the price of the share.
Even if we assume that the share is worth 100 monetary units, with a 1/100 tick size (the minimum price difference allowed), we shall have
a return $r$ larger than $1 \cdot 10^{-4}$ and much larger than $r_s$; therefore, it is safe to assume a vanishing risk-free interest 
rate. 

We shall focus on the price of an intra-day European call option assuming that the position is taken at a time $t$ coinciding with or
close to the beginning of the day and that it is closed at a later fixed time (the {\em maturity}) within
the same day, which we shall denote by $T_M$
(not to be confused with the epochs $T_i$, in general $T_M$ is {\em not} an epoch). Notice that the condition $t < T_M$ must
always be fulfilled.

Let $\widetilde{C}(S(T_M))$ represent the {\em pay-off} of a European call option at maturity. 
For instance, given the {\em strike price} $K$, the
pay-off of a plain-vanilla European option is $\widetilde{C}(S(T_M)) = \mathrm{max}(0,S(T_M) - K)$. 
Then, the option price $C(t)$ at a time $t < T_M$
is given by the discounted conditional expected value of the pay-off at maturity with respect to the e.m.m., that is
\begin{equation}
\label{optionprice}
C(t) = \mathrm{e}^{r(t-T_m)} \mathbb{E}_{\widetilde{\mathbb{S}}} (\widetilde{C}(S(T_M))|\mathcal{F}_t),
\end{equation}
where $r$ is the risk-free interest rate. In our case $r=0$, so that equation \eqref{optionprice} simplifies to
\begin{equation}
\label{optionpriceour}
C(t) = \mathbb{E}_{\widetilde{\mathbb{S}}} (\widetilde{C}(S(T_M))|\mathcal{F}_t).
\end{equation}
In order to evaluate equation \eqref{optionpriceour}, we consider two cases:
\begin{enumerate}
\item $t$ coincides with a renewal epoch;
\item $t$ does not coincide with a renewal epoch,
\end{enumerate}
with the second case being the only realistic one, but the first one is discussed in the recent literature as a
starting point for developing approximations as in \cite{cartea10}.

If the option price is evaluated from a renewal epoch, we can assume that $t=0$ without loss of generality and the
option price is given by the following integral
\begin{equation}
\label{optionpriceatrenewal}
C(0) = \mathbb{E}_{\widetilde{\mathbb{S}}} (\widetilde{C}(S(T_M))|\mathcal{F}_0) = \int_{0}^{\infty} 
\widetilde{C}(u) dF_{\widetilde{S}(T_M)} (u),
\end{equation}
where $F_{\widetilde{S}(T_M)} (u)$ is the cumulative distribution function of the random variable
$\widetilde{S} (T_M)$. In order to obtain this quantity, we can first define
\begin{equation}
\widetilde{S}_n = \prod_{i=1}^n \mathrm{e}^{Y_i - \log(\mathbb{E}(\mathrm{e}^{Y_i}))};
\end{equation}
$\widetilde{S}_n$ is the product of i.i.d. random variables and its cumulative distribution function is 
the $n$-fold Mellin convolution of $F_{\widetilde{Y}} (u)$, the common cumulative distribution function
of $\widetilde{Y}_i = Y_i - \log(\mathbb{E}(\mathrm{e}^{Y_i}))$; the Mellin transform is discussed in 
\cite{springer66} and in \cite{lomnicki67}. We shall write
\begin{equation}
\label{mellinconv}
F_{\widetilde{S}_n} (u) = F_{\widetilde{Y}}^{\star_{\mathcal{M}} n} (u).
\end{equation}
Since the number of trades from $0$ to $T_M$ can be an arbitrary integer, by purely probabilistic arguments, one can show that $F_{\widetilde{S}(T_M)} (u)$ is given by
\begin{equation}
\label{case1measure}
F_{\widetilde{S}(T_M)} (u) = \sum_{n=0}^\infty \mathbb{P} (N(T_M) = n) F_{\widetilde{Y}}^{\star_{\mathcal{M}} n} (u),
\end{equation}
as a consequence of the mutual independence of tick-by-tick log-returns and inter-trade durations. Notice that the zero-fold
Mellin convolution is a cumulative distribution function which is $0$ for $u=0$ and $1$ for $u>0$. To see that this is the case,
consider equation \eqref{optionpriceatrenewal} when it is known that $n=0$. Then $S(T_M) = S(0) = 1$ and the payoff is
$\widetilde{C}(S(0)) = \widetilde{C}(1)$. However, the probability $\mathbb{P}(N(T_M) = 0)$ 
of the event $N(T_M) = 0$ decreases with increasing $T_M$ and the contribution to the conditional expectation 
\eqref{optionpriceatrenewal} is $P(N(T_M) = 0) \widetilde{C} (1)$.
In order to use equation \eqref{case1measure}, we still need to compute the probabilities of the events $\{ N(T_M) = n \}$. This can be
again done by means of the indicator-function method. In fact, one has that
\begin{equation}
\label{countingevent}
\{ N(T_M) = n \} = \{T_n \leq T_M \} \cap \{T_{n+1} > T_M \}.
\end{equation}
Therefore, the following chain of equalities holds true
\begin{eqnarray}
\label{countingdistribution}
\mathbb{P} (N(T_M) = n) & = & \mathbb{P} (\{T_n \leq T_M \} \cap \{ T_{n+1} > T_M \}) = 
\mathbb{E} \left(I_{\{T_n \leq T_M \}} I_{\{ T_{n+1} > T_M \}} \right) \nonumber \\
& = & \mathbb{E} \left( I_{\{T_n \leq T_M \}} I_{\{ J_{n+1} > T_M - T_n \}} \right) = 
\int_{0}^{T_M} \int_{T_M - u}^\infty d F_J^{\star n} (u) d F_J (w) \nonumber \\
& = & \int_{0}^{T_M} (1 - F_J (T_M - u)) d F_J^{\star n} (u).
\end{eqnarray}

In the general case in which $t$ is a generic observation time not coinciding with a renewal epoch,
things become trickier, even if we are using a simplified and stylized model. At time $t$, both
the price $S(t)$ and the number of trades $N(t) = n_t$ are known. We can consider the random
variable $\Delta X(t,T_M) = X(T_M) - X(t) = \log(S(T_M)/S(t))$. If $S(t)$ is used as numeraire (that
is if we set $S(t) = 1$), Equation 
\eqref{optionpriceatrenewal} modifies to
\begin{equation}
\label{optionpricegeneral}
C(t) = \mathbb{E}_{\widetilde{\mathbb{S}}} (\widetilde{C}(S(T_M))|\mathcal{F}_t) = \int_{0}^{\infty} 
\widetilde{C}(u) dF^{n_t}_{\widetilde{S}(T_M)} (u),
\end{equation}
where the cumulative distribution function $F^{n_t}_{\widetilde{S}(T_M)} (u)$ is given by
\begin{equation}
\label{cumulativeconditional}
F^{n_t}_{\widetilde{S}(T_M)} (u) = \sum_{n=0}^\infty \mathbb{P}(N(T_M) - N(t) = n|N(t) = n_t) F_{\widetilde{Y}}^{\star_{\mathcal{M}} n} (u). 
\end{equation}
Again, as in the case of equation \eqref{optionpriceatrenewal}, this equation can be justified by purely probabilistic
arguments. However, one has to compute the conditional probability $\mathbb{P}(N(T_M) - N(t) = n|N(t) = n_t)$. 
As derived in \cite{kaizoji11}, this
is given by
\newpage
\begin{equation}
\label{countingcond}
\mathbb{P}(N(T_M) - N(t) = n|N(t) = n_t) = \int_{0}^{T_M - t} \mathbb{P} (N(T_M) - N(t+u) = n - 1) dF_{\mathcal{J}_{t,n_t}} (u),
\end{equation}
where $\mathbb{P} (N(T_M) - N(t+u) = n - 1)$ is given by equation \eqref{countingdistribution}  with $T_M$ replaced
by $T_M - (t +u)$ and 
$F_{\mathcal{J}_{t,n_t}} (u) = \mathbb{P}(\mathcal{J}_{t,n_t} \leq u)$ is the cumulative distribution function of 
the {\em residual life-time}
at time $t$ conditioned on the fact that there were $n_t$ trades up to time $t$ which we denote
by $\mathcal{J}_{t,n_t}$. The residual life time is the time interval from $t$ to the next renewal epoch $T_{N(t)+1}$.
As discussed in \cite{kaizoji11}, its distribution crucially depends on what
is known of the previous history. In our specific case, as anticipated above, it is meaningful to assume that we do know the total number of
trades up to time $t$, as this is usually public information. Before deriving $F_{\mathcal{J}_{t,n_t}} (u)$, it is important
to discuss the meaning of equation \eqref{countingcond}. The right-hand side contains the
probability of having $n -1$ trades between the renewal epoch $t+u$ and maturity $T_M$. Since the value $u$ of the residual
life time $\mathcal{J}_{t,n_t}$ is not known, this probability must be convolved with the probability of the event
$\{ \mathcal{J}_{t,n_t}= u\}$. It turns out that even the cumulative distribution function $F_{\mathcal{J}_{t,n_t}} (u)$ can be found by
direct elementary probabilistic tools without using Laplace-tranform methods. We can see that the event 
$\{\mathcal{J}_{t,n_t} \leq u\}$ can be described in term of a conditional event (see \cite{definetti95})
\begin{equation}
\label{conditionalevent}
\{ \mathcal{J}_{t,n_t} \leq u \} = \{ T_{n_t + 1} - t \leq u | N(t) = n_t \}.
\end{equation}
Equation \eqref{conditionalevent} can be written in terms of epochs using \eqref{countingevent}
\begin{equation}
\label{conditionaleventepochs}
\{ \mathcal{J}_{t,n_t} \leq u \} = \{ T_{n_t + 1} - t \leq u | \{T_{n_t} \leq t \} \cap \{T_{n_t+1} > t \} \}.
\end{equation}
One can now use the definition of conditional probability and the indicator-function method to compute
$F_{\mathcal{J}_{t,n_t}} (u)$ directly. First of all, one can write
\begin{eqnarray}
\label{residuallifetime}
F_{\mathcal{J}_{t,n_t}} (u) & = & \mathbb{P} (\mathcal{J}_{t,n_t} \leq u) = 
\mathbb{P}(T_{n_t + 1} - t \leq u | \{T_{n_t} \leq t \} \cap \{T_{n_t+1} > t \}) \nonumber \\
& = & \frac{\mathbb{P}(\{T_{n_t + 1} - t \leq u \} \cap \{T_{n_t} \leq t \} \cap \{T_{n_t+1} > t \})}{\mathbb{P}(\{T_{n_t} \leq t \} 
\cap \{T_{n_t+1} > t \})},
\end{eqnarray} 
and the denominator is already given by equation \eqref{countingdistribution}, meaning that one has
\begin{equation}
\label{denominator}
\mathbb{P}(\{T_{n_t} \leq t \} \cap \{T_{n_t+1} > t \}) = \int_{0}^{t} (1 - F_J (t - w)) d F_J^{\star n_t} (w).
\end{equation}
In order to compute the numerator, one can use the following equality between events
\begin{equation}
\label{numerator}
\{T_{n_t + 1} - t \leq u \} \cap \{T_{n_t} \leq t \} \cap \{T_{n_t+1} > t \} = \{T_{n_t} \leq t \} \cap
\{t - T_{n_t} < J_{n_t + 1} \leq t+ u - T_{n_t} \}, 
\end{equation}
and obtain that
\begin{eqnarray}
\label{numerator1}
\mathbb{P}(\{T_{n_t + 1} - t \leq u \} \cap \{T_{n_t} \leq t \} \cap \{T_{n_t+1} > t \})  & = & \nonumber \\
\mathbb{P}(\{T_{n_t} \leq t \} \cap \{t - T_{n_t} < J_{n_t + 1} \leq t + u - T_{n_t} \}) & = & \nonumber \\
\mathbb{E} \left( I_{\{T_{n_t} \leq t \}} I_{\{t - T_{n_t} < J_{n_t + 1} \leq t + u - T_{n_t} \}} \right) & = &
\int_{0}^t \int_{t - w}^{u + t - w} dF_{T_{n_t}} (w) dF_{J} (v)  = \nonumber \\
\int_{0}^t \int_{t - w}^{u + t - w} 
dF^{\star n_t}_{J} (w) dF_{J} (v)  & = & \nonumber \\ 
\int_0^t (F_J (u+t-w) - F_J (t-w)) dF_J^{\star n_t} (w).
\end{eqnarray}
Combining equations \eqref{denominator} and \eqref{numerator1}, one finally gets from equation \eqref{residuallifetime}
\begin{equation}
\label{residuallifetimefinal}
F_{\mathcal{J}_{t,n_t}} (u) = \frac{\int_0^t (F_J (u+t-w) - F_J (t-w)) dF_J^{\star n_t} (w)}{\int_{0}^{t} 
(1 - F_J (t - w)) d F_J^{\star n_t} (w)}.
\end{equation}
Equation \eqref{residuallifetimefinal} is the last ingredient needed to determine the option price in the
general case \eqref{optionpricegeneral}. Finally, note that equation \eqref{optionpricegeneral} yields 
equation \eqref{mertonformula} when $J \sim \exp(\lambda)$ and $Y \sim N(\mu,\sigma^2)$ (see 
\cite{baleanu12}, chapter 7).

\section{Discussion and Outlook}
\label{concl}

Equation \eqref{optionpricegeneral} is our main results and gives a martigale price for intraday European options 
when assuming the parsimonious model of section \ref{tick}; we have been able to explicitly derive all the terms in that equation 
by repeated application of the indicator-function method. 
Even if such an equation may seem cumbersome, we already showed that it
can be used in practice in \cite{kaizoji11}. A more detailed numerical analysis will be the subject 
of future research.

However, some assumptions in section \ref{tick} are unrealistic even if they ensure analytical tractability. 
For example, it is assumed that the durations
$\{J_i\}_{i=1}^\infty$ and the tick-by-tick log-returns
$\{Y_i\}_{i=1}^\infty$ are i.i.d. random variables and that they are mutually independent. In \cite{engle97}. \cite{engle98},
\cite{raberto02} and 
\cite{meerschaert06}, as well as in many other empirical papers on financial econometrics (see \cite{campbell96}), it is shown that
this is not the case. There is heteroscedasticity and there is dependence between the activity and the volatility. Suitable
mixture models based on the compound Poisson processes can take all that into account as discussed in \cite{scalas07}, but models
using heteroscedastic procesess subordinated to Hawkes processes could be a viable alternative as well (see \cite{munitoke11}
and references therein).

This paper is the crowning achievement of an activity on modelling ultra-high frequency financial data by means of 
continuous time random walks that started back in 1998. Continuous time random walks is the name that physicists use for compound renewal processes, even
if some authors reserve this name to more general processes with finite or infinite memory subordinated to counting processes. 
As briefly discussed above, these processes allow the derivation of many non-trivial analytical results, 
but they are not general enough to take into account all the features of high-frequency financial data. In 1998, the idea
was to use these processes for intra-day option pricing, but only in 2011, with the results published in \cite{kaizoji11}, it
became possible to present martingale option pricing in the simple way based on renewal theory outlined in this paper.

\section*{Acknowledgement}
This work was partially funded by MIUR Italian grant PRIN 2009 on {\em Finitary and non-Finitary Probabilistic Methods in Economics}
within the project {\em The Growth of Firms and Countries: Distributional Properties and Economic Determinants}. One of the
authors (E.S.) gratefully acknowledges discussion with Giacomo Bormetti, Giacomo Livan and Fabio Rapallo.

\bibliographystyle{EconEJ}
\bibliography{SParxiv}

\end{document}